# The Effects of Spectrograph Slit Modes on the Accuracy of Stellar Radial Velocity Measurement and Atmospheric Parameter Estimation

WANG FengFei[1, 2, 3], ZHANG HaoTong[1, 2*], LUO A-Li[1, 2], BAI ZhongRui[1, 2, 3], DU Bing[2, 3], ZHAO YongHeng[1, 2]

[1] National Astronomical Observatories, Chinese Academy of Sciences, 20A Datun Road, Chaoyang District, Beijing 100012, China
[2] Key Laboratory of Optical Astronomy, National Astronomical Observatories, Chinese Academy of Sciences, 20A Datun Road, Chaoyang District, Beijing 100012, China
[3] University of Chinese Academy of Sciences, No.19A Yuquan Road, Beijing 100049, China

Spectrograph slit is conventionally used to enhance the spectral resolution and manage how much light can be allowed to enter spectrograph. The narrow slit provides a higher resolution but sacrifices efficiency of spectrograph and results in a low signal to noise ratio (S/N) spectra product. We take GuoShouJing telescope as an example and carry out a series of experiments to study how its 2/3 slit mode affects the precision of stellar radial velocity measurement and atmosphere parameters estimate. By transforming the resolution and adding a Gaussian White Noise to the extremely high quality spectra from the Sloan Digital Sky Survey, we generate synthetic stellar spectra of various brightness with different S/Ns. Comparing the measurements on these noise added spectra with the original high quality ones, we summarize the influences of the 2/3 slit mode on the measurement accuracy of stellar radial velocity and atmospheric parameters.

Citation:

In traditional long slit spectrum observation, a slit is usually put in the entrance pupil of the spectrograph to block extra light from the sky and other objects. For a certain grating, the resolution of the spectrograph is determined by the width of the slit, and thus the slit width is adjusted according to the seeing conditions for most of the object light to enter the spectrograph and maintain the best resolution simultaneously. Generally, no slit is adopted for fiber-spectrum as the input light is constrained by the fiber aperture and no more blocks are needed. The resolution is determined by the diameter of the fiber. GuoShouJing telescope (a.k.a. LAMOST) [1][2] is a reflecting Schmidt telescope specialized for conducting spectroscopic surveys, located at Xinglong Station, Hebei Province, China. Its ability to obtain spectra of 4,000 objects simultaneously makes it the largest and most efficient telescope of China. The telescope is planned to conduct a 5-year spectroscopic survey of 5 million Milky Way stars, as well as millions of galaxies. The 16 spectrographs used in LAMOST can be run under low-resolution mode with a full slit or a 2/3 slit, resulting in the resolution of 1200 or 1800 respectively. In full slit mode, the star light is guided directly into the spectrograph entrance pupil by the fibers, the ends of the fibers are aligned in a line by clips and there is no actual slit inside. The PSF (point-spread function) of the spectrograph is determined by the diameter of the fiber which is $320 \mu$ m. For the 2/3 slit mode, a slit with 2/3 width of the fiber is installed in front of the clips to narrow down the light beams from the fibers. For spectra with high S/Ns, higher resolution due to the 2/3 slit brings higher accuracy of measurements of stellar radial velocities and atmosphere parameters since the full width at the half maximum (FWHM) of spectral lines becomes narrower, making it easier to determinate the radial velocity of stellar spectra and estimate stellar atmosphere parameters, including effective temperature (Teff), surface gravity (Log$g$), and metallicity ([Fe/H]) as more subtle features of spectra are resolved. On the other hand, the 2/3 slit mode provides higher resolution at the cost of losing more light, which means lower S/N spectra product. Therefore, one needs to choose the slit mode based on the specific scientific goal. For example, galaxy observers may give priority to higher S/N while stellar researchers prefer higher resolution to view more features of stars in the Milky Way. Balancing the LAMOST different projects including the LAMOST ExtraGAlactic Survey (LEGAS) [2], and the LAMOST Experiment for Galactic Understanding and Exploration (LEGUE) [3] survey of Milky Way, the scientific committee of LAMOST telescope decided to use slits with 2/3 fixed width for LAMOST surveys, which meets the goals of two surveys.

*Corresponding author (email: zht@lamost.org)

We know how slit modes affect the efficiency and resolution, but with no quantitative study on how much they affect the accuracy of stellar radial velocity and atmosphere parameters determination. In this paper, we study this kind of influence using more than 3,000 high quality stellar spectra from the Sloan Digital Sky Survey (SDSS) [4] as our sample. We first calculated the theoretical counts and noise level for different star brightness levels, and then degraded those high quality spectra accordingly. Next we measured the radial velocities and parameters of these noise added spectra by using our stellar parameter pipeline. We repeat the above steps on full slit and 2/3 slit mode respectively to examine how much the different slit modes affect the measurement precision.

**Table 1** The signal-to-noise per pixel for stars with different magnitudes under specific slit and efficiency modes.

| $g$-band | | | | |
|---|---|---|---|---|
| Mag / Mode | 16 | 17 | 18 | 19 |
| Slit1 | 51 | 29 | 15 | 7 |
| 2/3slit(78%) | 45 | 25 | 13 | 6 |
| 2/3slit(60%) | 39 | 22 | 11 | 5 |
| $i$-band | | | | |
| Mag / Mode | 16 | 17 | 18 | 19 |
| Slit1 | 51 | 29 | 15 | 7 |
| 2/3slit(78%) | 45 | 25 | 13 | 6 |
| 2/3slit(60%) | 39 | 22 | 11 | 5 |

Theoretically, the efficiency of 2/3 slit mode is supposed to be 78% of the full slit. In practice, this proportion varies from 60% to 78% based on a series of inner experiments on spectrograph slit. In this paper, we will calculate stellar parameters with modes of full slit (hereafter, slit1), 2/3 slit with 60% efficiency (hereafter, 2/3slit60%), and 2/3 slit with 78% efficiency of full slit (hereafter, 2/3slit78%). Assuming that LAMOST has an effective aperture of 4.5 meters, and observing a plate with exposure time 1800 seconds, we calculated the theoretical S/N per pixel of a stellar object at the wavelength of 4770 angstrom (center of SDSS $g$-band) and 7625 angstrom (center of SDSS $i$-band) with magnitudes of 16, 17, 18, and 19 respectively. Corresponding S/N values of different slit modes are listed in Table 1.

# 1 Sample and Synthetic Spectra

The SDSS, with similar wavelength range coverage and resolution to LAMOST, provides millions stellar spectra product. We select stellar spectra with best qualities from SDSS data release 7 (DR7) with S/Ns over 80. The specific query command used for SDSS CasJobs tool [5] is "select plate, mjd, fiberid, sptypea, sna, g0, ELODIErv from sppparams where sna>80 and g0<16 and g0>12 and abs(ELODIErv)<200" and we obtain 3,207 objects. These spectra are good enough to be considered as signal without noise. The spectral types of these objects include A, F, G, and K type stars. The distribution of $g$ magnitude of this set, showed in Figure 1, is concentrated at 15 ~ 16 and $g$ –$r$ color of 0.1 to 0.7. The gap between 0.4 and 0.5 shows the Sloan Extension for Galactic Exploration and Understanding (SEGUE)[5] has a selection effect here and there are not many super high S/N stellar spectra in this region.

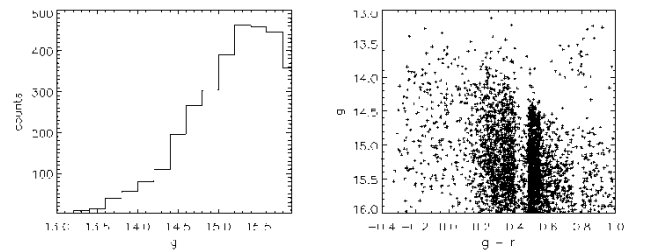

**Figure 1** The distribution of magnitudes of g and the color-magnitude diagram of our sample. The gap between 0.4 and 0.5 is due to the selection effect of SEGUE.

In order to generate LAMOST full slit and 2/3 slit modes spectra, we need to change the SDSS resolution to corresponding LAMOST resolutions. We determine the resolution of SDSS spectra as 2100 and LAMOST as 1800 at 5500 angstrom by measuring the FWHM of sky emission lines. Employing the de-resolution equation as below, we degraded the SDSS spectra to LAMOST resolution by convolving a Gaussian kernel with sigma ($\sigma$) equal to 0.67

$$\sigma = \sqrt{\left(\frac{5500}{1800}\right)^2 - \left(\frac{5500}{2100}\right)^2} \Big/ 2.35 = 0.67$$

Then we have obtained the simulative LAMOST 2/3 slit mode high quality spectra with a resolution of 1800. If we remove the slit (then it is full slit), the resolution is supposed to be 1200 in theory. So we use the following equation to generate the LAMOST full slit mode high quality spectra with a resolution of 1200.

$$\sigma = \sqrt{\left(\frac{5500}{1200}\right)^2 - \left(\frac{5500}{1800}\right)^2} \Big/ 2.35 = 1.45$$

The next step for synthesizing spectrum is to generate different S/N spectra to simulate LAMOST observed stars of different magnitudes in different slit modes. Taking the magnitude of 16 as an example, based on Table 1, the S/N is 51 for slit1, 45 for 2/3slit78%, and 39 for 2/3slit60%. A Gaussian White Noise is added to all the high S/N spectra from previous step to generate spectra with specific S/N. Figure 2 shows an example of a g-band magnitude of 16, with the red line as the original spectrum without noise. The blue one is the slit1 (R = 1200) spectra with an S/N of 51; the green one is the 2/3slit78% (R = 1800) spectra with an S/N of 45 and the black one is the 2/3slit60% (R = 1800) spectra with an S/N of 39.

In this way, we produce 12 groups' spectra with g-band magnitude in [16, 17, 18, 19] under a slit mode of slit1, 2/3slit78%, and 2/3slit60%.

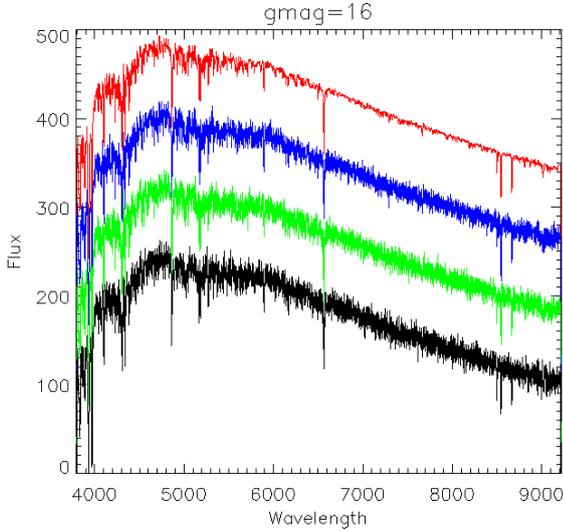

**Figure 2** An example of spectrum with a g-band magnitude of 16. A Gaussian White Noise is added on the original high quality spectra to generate different slit modes spectra. The red line is the original spectrum without noise added. The blue one is the slit1 (R = 1200) spectra with an S/N of 51; the green one is the 2/3slit78% (R = 1800) spectra with an S/N of 45 and the black one is the 2/3slit60% (R = 1800) spectra with an S/N of 39.

## 2 Influence on Measurements of Radial Velocity

The 1D-pipeline software of LAMOST [1][6] employs the same method as SDSS [5] to measure the radial velocities of stellar objects. We estimate our stellar radial velocities by performing a "best-match" method that compares the observed spectra with externally measured templates --- the ELODIE library of high-resolution spectra [7]. This method, following the spirit of Glazebrook et al.[8], is also adopted by specBS pipeline [9], which is written for SDSS spectral analysis by D. Schlegel et al.. The spectra in the ELODIE library were obtained with the ELODIE spectrograph of 1.93 m telescope at the Observatories de Haute-Provence. The wavelength coverage is from 4000 to 6800 angstrom. We employ 1962 spectra of 1388 stars with a resolution R ~ 10,000, which are publicly available as part of the ELODIE 3.1 release [7]. The resolution of these spectra is degraded to 1800 to match the LAMOST spectra by convolving with a Gaussian kernel. Most of the spectra have quite high S/Ns, and are accompanied with estimated stellar parameters from the literature.

We measure the radial velocities (RV) of 3,207 stellar spectra with high quality; the difference between our RV and those from SDSS Online Database is plotted in Figure 3. As we can see, the scatter of their differences is 0.6km/s, which is absolutely small, proving that our pipeline is suitable for these spectra. Notice that a -7.3km/s systematic offset is added on the ELODIE radial velocities of all SDSS stellar spectra for zero-point correction [10], which is coincident with our results in Figure 2. After removing a few points which have a relatively large difference (>10km/s), the count of our data sample is reduced to 3,175.

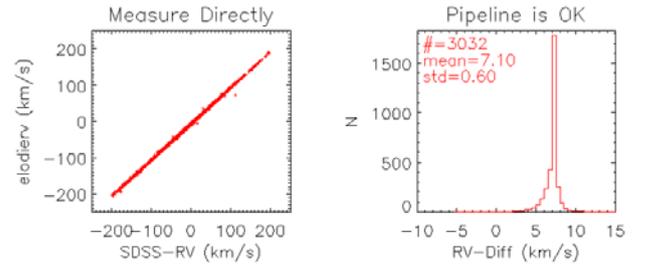

**Figure 3** By measuring the radial velocities by 1D-pipeline, compared with SDSS results, a tiny scatter proves our method and the results are reliable. The "mean", is 7.10 km/s, is the mean value of the differences of our RV and SDSS RV and the "std" is the standard deviation of the differences with a value of 0.6 km/s, which is considered a very slight scatter.

The radial velocities measured from spectra without noise added on (RV0, here after) become the prototype of other 12 groups dataset mentioned in section 1 to assess the accuracy of measurement on those noise added spectra. The radial velocities from spectra of each group are compared with RV0. The mean value (the systematic offset) and the standard deviation (std.) of the differences (the statistical 1-sgima error) are summarized in Table 2. The std. of the differences between noise added spectra radial velocity and RV0 are plotted in Figure 4 in order to be more intuitive.

According to the "std." from Table 2 and Figure 3, the 2/3 slit spectrograph mode improves the accuracy of radial velocity measurement slightly even under the lower 60% efficiency (except for the magnitude 19 spectra). An improvement of 0.5 ~ 1.5km/s indicates that the 2/3 slit mode plays a positive role in the stellar radial velocity measurement, especially for bright stars. However, a 20.44km/s error for the mode of 2/3slit(60%) with a g-band magnitude of 19, where the S/N is only about 5, indicates that higher resolution does not always obtain high accuracy radial velocity when the S/N is low.

**Table 2** The mean values and the "std" of radial velocity differences between 12 groups noise added spectra and the original high quality ones.

| std: 1-sigma error (km/s) | | | | |
|---|---|---|---|---|
| Mag-g<br>Mode | 16 | 17 | 18 | 19 |
| Slit1 | 2.92 | 4.72 | 9.00 | 18.80 |
| 2/3slit(78%) | 2.44 | 3.95 | 7.56 | 16.50 |
| 2/3slit(60%) | 2.74 | 4.48 | 8.69 | 20.44 |
| mean: systematic error (km/s) | | | | |
| Mag-g<br>Mode | 16 | 17 | 18 | 19 |
| Slit1 | -0.52 | -0.43 | -0.06 | 1.60 |
| 2/3slit(78%) | -0.26 | -0.17 | 0.19 | 1.93 |
| 2/3slit(60%) | -0.27 | -0.15 | 1.69 | 2.11 |

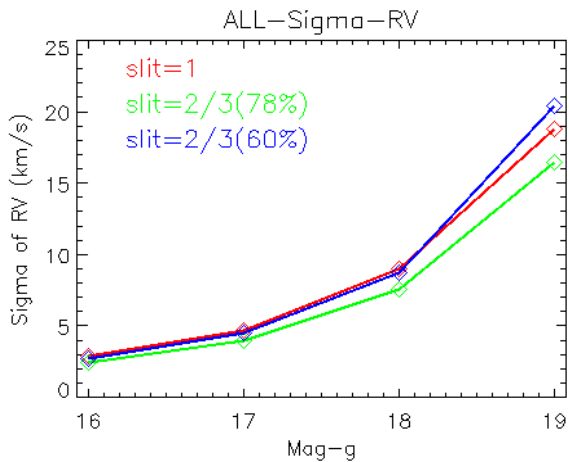

**Figure 4** The 1-sigma statistical differences obtained from comparing the radial velocities between RV0 and 12 groups spectra. Every square point is the 1-sigma scatter for each group, which contains specific magnitude and S/N spectra. The red line is the error of full slit with R ~ 1200. The green line is for the 2/3slit(78%) and the blue is 2/3slit(60%), both of which have a resolution of 1800.

For more inspection on the radial velocity accuracy variation, we separate our samples by spectral type as A, F, G, K type subsets. We compare the radial velocities precision of different spectral types and display the statistic differences in Figure 5. For the FGK stars, the accuracy of RV measure for 2/3slit60% is similar to the full slit mode, and it's only a bit better (up to 4km/s) for 2/3slit78%. In the case of A type star, full slit is better than 2/3slit60%, similar to 2/3slit78%, or even better at an magnitude of 19. 2/3slit60% has about 15km/s more scatter than the full slit mode at the magnitude 19. Since the efficiency of the slit varies between 60% and 78%, we conclude that 2/3 slit may improve the RV accuracy slightly for FGK type stars, but does not affect the A type star

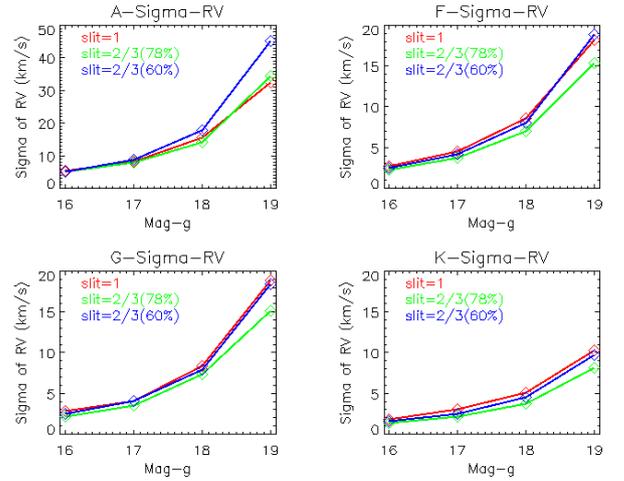

**Figure 5** The 1-sigma statistical differences of different spectral type stars. The top-left is for A stars, the top-right is for F stars, the bottom-left is for G stars, and the bottom-right is for K stars. The lines here are the same as in Figure 4. 2/3 slit improves the G and K stars radial velocity accuracy while makeing the A and F worse.

# 3 Effects on the Estimate of Stellar Atmosphere Parameters

The SDSS employs a software package, the Sloan Extension for Galactic Exploration and Understanding (SEGUE) Stellar Parameter Pipeline (SSPP), to estimate the fundamental stellar atmospheric parameters (effective temperature, surface gravity and metallicity) for AFGK-type stars using multiple techniques based on medium-resolution spectroscopy and *ugriz* photometry [10]. A number of tests were carried out to assess the performance of the SSPP. They sought to validate the radial velocities and stellar parameters determined by the SSPP by comparing selected member stars of three globular clusters (M 15, M 13, and M 2) and two open clusters (NGC 2420 and M 67). Besides, 150 high-resolution spectra of SDSS-I/SEGUE stars were used to calibrate and validate the radial velocities and the atmospheric parameters adopted by the SSPP [11].

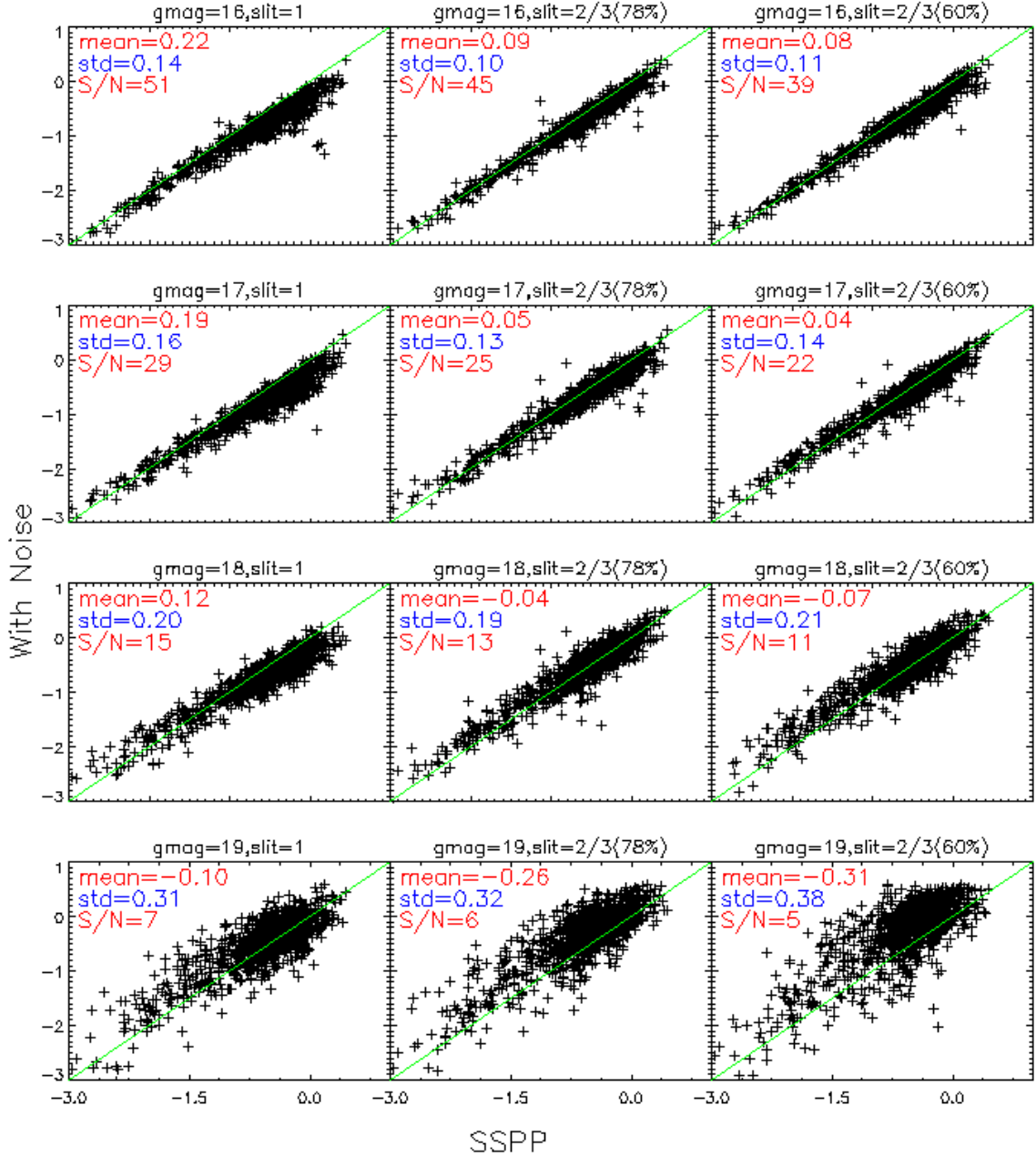

**Figure 6** Comparison of FeH0 with 12 groups spectra under full slit and 2/3 slit mode with different S/N. The horizontal axis is the [Fe/H] from original high S/N spectra and the vertical axis is the ones with noises. The "mean" marked in the diagram shows the mean of the differences (the systematic error) between the original spectra and the noise-added spectra. The "std" here is the standard deviation of the differences (the 1-sigma statistical error).

In this paper, we focus on studying how much 2/3 slit influences [Fe/H] and *Teff* estimate accuracy. We divided the sample into four groups according to their spectral types, A F G K, each containing about 300 stars. Then those stars were degenerated based on slit mode and magnitude as described in sections1 & 2, with finally 12 groups of data samples in different spectral types and slit modes. We employ ki13, NGS, ANRR, ANSR, HA24, and HD24 methods of SSPP for *Teff* estimate while only choosing ki13, NGS, ANRR, and ANSR for [Fe/H] estimate. Then a combination of these methods is used on each group spectra respectively. We also apply those methods to the original high quality spectra for comparison.

To further investigate how the 2/3 slit affects the parameter accuracy, we use the high S/N SDSS spectra as a standard to compare the difference between the high S/N and the degenerated spectra. We define the effective temperature and metallicity for high S/N results as Teff0 and FeH0 respectively. Figure 6 shows the respective [Fe/H] differences of full slit and 2/3slit (60% & 78%) compared with FeH0. The standard deviation of the differences ("std." in Figure 6) is larger with lower S/Ns. The differences between noises added spectra and FeH0 for different slit modes show similar level in dispersion. The 2/3 slit mode is a little bit smaller than the full slit mode except in magnitude 19, while no obvious improvement is found for faint stars.

Figure 6 also indicates that there is a systematic offset with S/N for each slit mode. Spectra with higher S/N show an underestimate of [Fe/H], while with the increase of noise, the systematic offset turns positive, and [Fe/H] becomes overestimated. This systematic effect with noise for different methods had already been noticed by SSPP [10], and can be corrected with prior knowledge from this kind of simulation. The standard deviation of the differences of each spectral type star are plotted in Figure 7. The influence on metal-rich K type stars is very small while metallicity error shows a relatively larger scatter on the stars with less metal lines, such as A and F type stars.

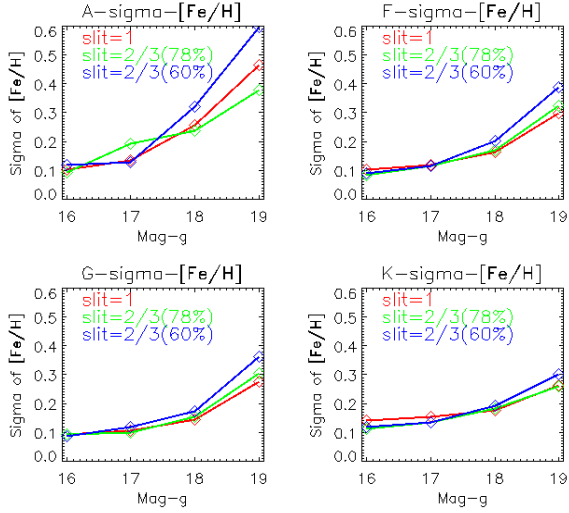

**Figure 7** The statistic metallicity error of different spectral types The top left is for A stars, the top right is for F stars, the lower-left is for G stars, and the lower right is for K stars. The lines here are the same as in Figure 3.

We employ the SSPP methods to estimate the effective temperature and compare the results with Teff0 in the same way as we did for metallicity. According to Figure 8, in most cases except for 2/3slit78% at magnitude 16, full slit has lower scatter than 2/3slit in temperature determination. The 2/3 slit decreases the accuracy of *Teff* estimation 2 ~ 60K. We also check this effect on different spectral types in Figure 9 and draw the same conclusion, indicating that 2/3 slit does not improve the *Teff* estimation for different spectral types. Figure 8 also indicates that there is a system underestimation of temperature for all slit modes when S/N decreases. According to the experiments by Lee et al.[10], the temperature determination is sensitive to S/N for some of the methods in the SSPP, especially ANNSR, NGS1 and HA24 methods used here. So we conclude that the system change of *Teff* determination with S/N is not caused by the slit mode, but it can be corrected once the system pattern is known and won't affect our conclusion that the 2/3 slit won't improve the accuracy of Teff determination.

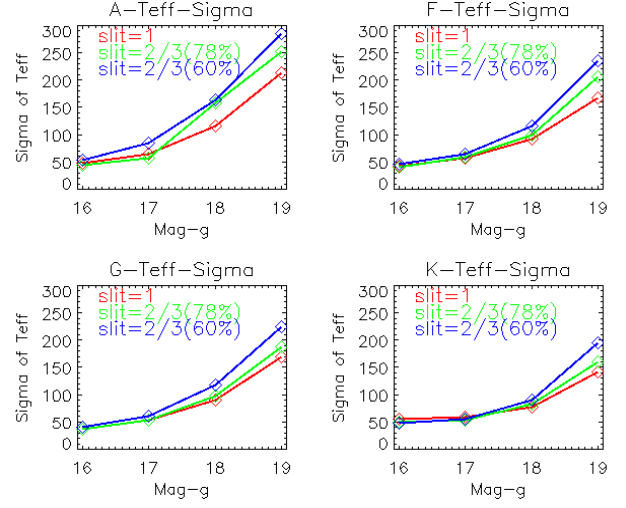

**Figure 9** The 1-sigma statistical error for different spectral types obtained from comparing the effective temperatures between Teff0 and 12 groups spectra.

## 5 Conclusions

In this paper, we generate synthetic stellar spectra of various brightness by adding a Gaussian White Noise to simulate the 2/3 slit (60% and 78% efficiency) and full slit mode of LAMOST spectrograph for the purpose of assessing how much 2/3 slit improves the measurement of stellar radial velocity and atmosphere parameters for a given magnitude. A series of experiments reveal the 2/3 slit mode influences the accuracy of radial velocity measurements with a slight improvement of 0.5 ~ 1.5km/s except for the A type star which has less lines in their spectra than later types. In terms of stellar atmosphere parameters, we estimate the metallicity and effective temperature of stellar spectra grouped by spectral type. The 2/3slit78% mode improves the [Fe/H] measurement accuracy of bright stars with high S/N while no obvious improvement is proved for faint stars compared with the full slit. Besides, the 2/3slit60% mode is similar to full slit mode. In addition, the [Fe/H] of the metal-rich K type stars causes a small error and the effect from 2/3 slit is not obvious while the accuracy of relatively metal-poor stars, such as A, F type stars, is reduced because of low S/N of 2/3 slit. The experiments on effective temperature turn in a consistent performance that 2/3 slit mode decreases the accuracy of *Teff* estimation for various brightness stars with an order of 2 ~ 150K compared with the full slit mode. The systematic difference of *Teff* estimation accords with the experiments on parameter sensitivities to S/N by Lee et al. [10].

Since the efficiency of 2/3 slit changes between 60% and 78%, the accuracy of determined stellar parameter should vary between those values in simulation. Then we conclude that the 2/3 slit improves RV and [Fe/H] measurement accuracy, especially for the bright stars, but with a drawback for $T_{eff}$ estimate since it decreases the efficiency of spectrograph and gains low S/N spectra.

*This work was supported by the National Natural Science Foundation of China (Grant Nos. 11203045).*

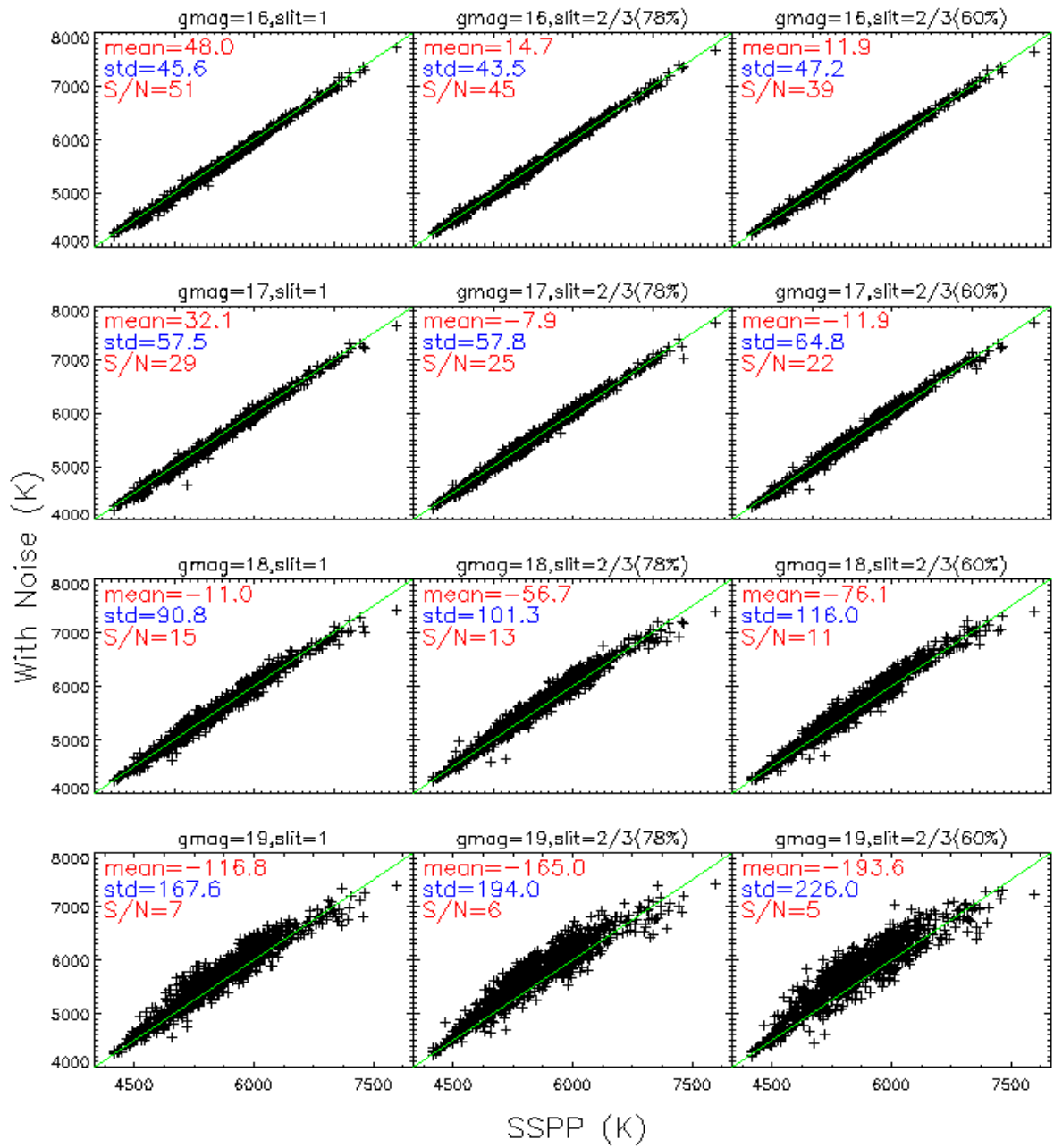

**Figure 8**   Comparing Teff0 with 12 groups spectra under full slit and 2/3 slit mode with different S/N.